\documentclass[singlecolumn,amsmath,amssymb]{revtex4}

\usepackage{graphicx}
\usepackage{bm}

\begin{document}

\title{Pressure-induced reconstitution of Fermi surfaces and spin fluctuations  in S-substituted FeSe}
\author{T. Kuwayama,$^1$ K. Matsuura, $^2$\footnote{present address: Research Center for Advanced Science and Technology, University of Tokyo, 4-6-1 Komaba, Meguro-ku, Tokyo 153-8904, Japan} J. Gouchi,$^3$ Y. Yamakawa,$^4$ Y. Mizukami,$^2$ S. Kasahara,$^5$\footnote{present address: Department of Physics, Okayama University, Okayama 700-8530, Japan} Y. Matsuda,$^5$ T. Shibauchi,$^2$ H. Kontani,$^4$ Y. Uwatoko,$^3$ and N. Fujiwara$^1$\footnote{Corresponding author: fujiwara.naoki.7e@kyoto-u.ac.jp}}
\affiliation{$^1$Graduate School of Human and Environmental Studies, Kyoto University, Yoshida-Nihonmatsu-cyo, Sakyo-ku, Kyoto 606-8501, Japan\\
$^2$Graduate School of Frontier Sciences, University of Tokyo, 5-1-5 Kashiwanoha, Kashiwa, Chiba 277-8581, Japan\\
$^3$Institute for Solid State Physics, University of Tokyo, 5-1-5 Kashiwanoha, Kashiwa, Chiba 277-8581, Japan\ \\
$^4$Department of Physics, Nagoya University, Furo-cho, Nagoya 464-8602, Japan \ \\
$^5$Division of Physics and Astronomy, Graduate School of Science, Kyoto University, Kitashirakawa Oiwake-cho, Sakyo-ku, Kyoto 606-8502, Japan}


\begin{abstract}
\bf FeSe is a unique high-$T_c$ iron-based superconductor in which nematicity, superconductivity, and magnetism are entangled with each other in the $P-T$ phase diagram. We performed $^{77}$Se-nuclear magnetic resonance measurements under pressures of up to 3.9 GPa on 12\% S-substituted FeSe, in which the complex overlap between the nematicity and magnetism are resolved. A pressure-induced Lifshitz transition was observed at 1.0 GPa as an anomaly of the density of states and as double superconducting (SC) domes accompanied by different types of antiferromagnetic (AF) fluctuations. The low-$T_{\rm c}$ SC dome below 1 GPa is accompanied by strong AF fluctuations, whereas the high-$T_{\rm c}$ SC dome develops above 1 GPa, where AF fluctuations are fairly weak. These results suggest the importance of the $d_{xy}$ orbital and its intra-orbital coupling for the high-$T_{\rm c}$ superconductivity.
\end{abstract}

\maketitle

FeSe has unusual features among high-$T_c$ iron-based superconductors \cite{Matsuda} because its superconductivity emerges without magnetism in the nematic phase where four-fold rotational symmetry breaks \cite{Fernandes2014, Shimojima2014, Shimojima2019}. The absence of magnetism originates from characteristic unconnected Fermi surfaces (see the unfolded Fermi surfaces in the left panel of Fig. 1a): small hole pockets at point $\Gamma$, $\textbf{\emph{k}}=(0,0)$, and anisotropic electron pockets at point $X$, $\textbf{\emph{k}}=(\pi,0)$ or $(0,\pi)$, which are caused by the splitting of the $d_{xz}$ and $d_{yz}$ orbitals \cite{Watson2015a, Watson2017, Kushnirenko2018, Coldea2018, Skornyakov2018, Fanfarillo2018}. The orbital configuration at ambient pressure (see the left panel of Fig. 1a) reduces the likelihood of nesting between electron and hole pockets with the same orbital, leading to the absence of magnetism. The importance of orbital selectivity in Cooper pairing for the superconducting (SC) state in the nematic phase has been suggested\cite{Sprau}.

Upon pressure application, FeSe undergoes an antiferromagnetic (AF) order instead of the nematic order. The AF order is accompanied by an enhancement in $T_{\rm c}$: the $T_{\rm c}$ of 9 K at ambient pressure increases to 38 K at pressures above 6 GPa \cite{Sun2016a}.
Nematicity, superconductivity, and magnetism are entangled with each other in the pressure versus temperature ($P$-$T$) phase diagram. This makes it extremely difficult to understand the nature of this system, although S substitution resolves the complex overlap between the nematic and AF phases, and rich-S substitution induces the nematic critical phenomenon \cite{Matsuura2017, Xiang2017, Holenstein2019}.
Furthermore, experimental difficulties are faced in observing Fermi surfaces under pressure-restricted experimental approaches. In fact, direct observations via angle-resolved photoemission spectroscopy (ARPES) or scanning tunneling microscopy (STM) have not been reported so far. To date, only a few experimental results have been reported. In particular, the Hall coefficient changes sign from minus to plus upon pressure application \cite{Sun2017}, and the band masses for several orbitals gradually change around the nematic critical point (0.58 GPa) \cite{Col2019}. The presence of a stripe-type AF order with $\textbf{\emph{Q}}$=($\pi$, 0) or (0, $\pi$) has been suggested from the results of nuclear magnetic resonance (NMR) measurements \cite{Wang2016, Wiecki2017a}. NMR measurements on 12\% S-substituted FeSe have revealed that the characteristics of low-energy magnetic fluctuations change at 1 GPa \cite{Kuwa}, which is indicative of the reconstitution of Fermi surfaces as well as the band mass change.

\begin{figure*}
\includegraphics[clip, width=1\columnwidth]{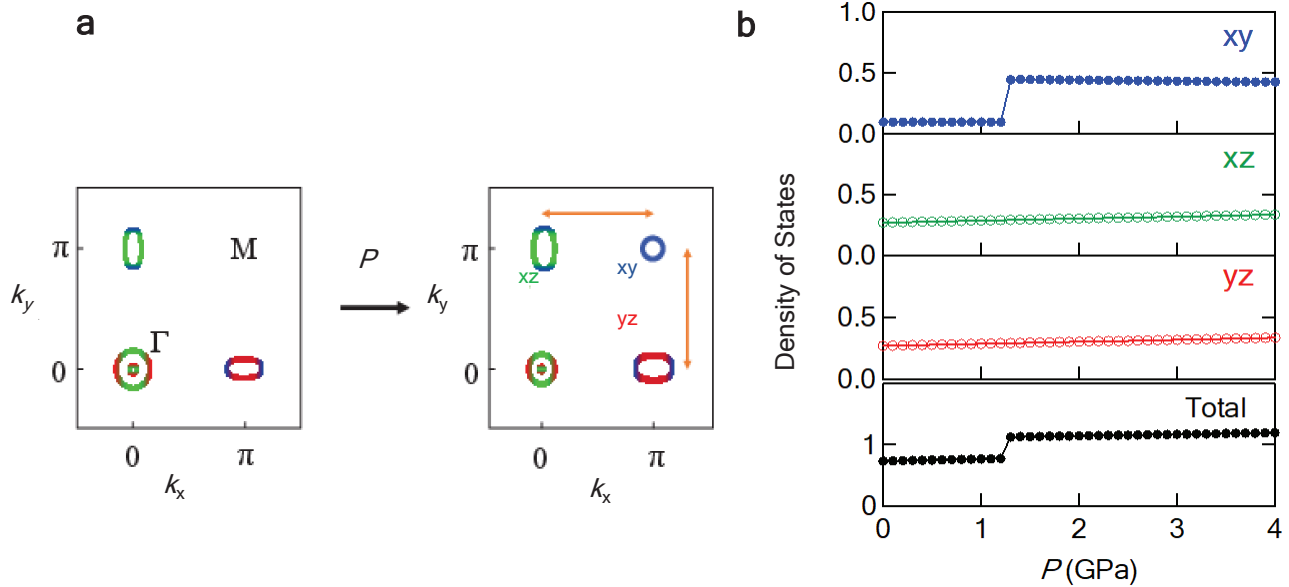}
\caption{\label{fig:epsart} {Contribution of each orbital to unfolded Fermi surfaces and the density of states (DOS). } {\bf a,} Schematic Fermi surfaces of pure FeSe in a tetragonal phase theoretically derived at ambient pressure \cite{Watson2015a, Watson2017, Kushnirenko2018, Coldea2018, Skornyakov2018, Fanfarillo2018} and high pressure \cite{Yamakawa2017a, Sun2017}. The Fermi surfaces colored in green, red, and blue represent $d_{xz}$, $d_{yz}$, and $d_{xy}$ orbitals, respectively. In a high-pressure regime, the antiferromagnetic (AF) order can be induced owing to the nesting of $d_{xy}$ orbitals between points M and X, as indicated by arrows. {\bf b,} The DOS originating
from $d_{xy}$, $d_{xz}$, and $d_{yz}$ orbitals and the total DOS calculated theoretically based on the crystal structure of 10\% S-substituted FeSe.  }
\end{figure*}

To explain the appearance of the AF order under pressure, a theoretical model has been proposed. In this model, another hole pocket emerges with increasing pressure at point $M$, $\textbf{\emph{k}}=(\pi,\pi)$, causing a better nesting with the electron pocket at point $X$ (see the unfolded Fermi surfaces in the right panel of Fig. 1a \cite{Yamakawa2017a}). The better nesting between points $X$ and $M$ with the same orbital can induce a stripe-type magnetic order. When the hole pocket emerges at point $M$, the shapes of pockets at points $\Gamma$ and $X$ are qualitatively similar to those at ambient pressure, although their size changes monotonically with increasing pressure \cite{Yamakawa2017a}.  The emergence of the $d_{xy}$ hole pocket also enhances the density of states (DOS) (see Fig. 1b), as will be described in detail later. The pressure-induced reconstitution of Fermi surfaces can change the Cooper pairing, leading to an SC-SC phase transition from the SC state under the nematic order to a higher-$T_{\rm c}$ state (see Fig. 4b). The higher-$T_{\rm c}$ state would provide an intriguing stage for the superconductivity mechanism common to iron-based superconductors with a high $T_{\rm c}$.
However, such a theoretically predicted Lifshitz transition has not been reported so far because of the entangled $P-T$ phase diagram and the experimental difficulties faced in observing Fermi surfaces under pressure.

In this study, we conducted $^{77}$Se(I=1/2)-NMR measurements under pressure, focusing on 12\% S-substituted FeSe, where the overlap of nematicity and magnetism is absent in the intermediate-pressure regime between 1 and 4 GPa \cite{Matsuura2017, Xiang2017}. Based on the results, we suggest that the theoretically predicted Lifshitz transition is observed as an anomaly of the DOS and as double SC domes accompanied by different types of AF fluctuations.

Typical NMR spectra corresponding to the nematic and magnetic orders are shown in Fig. 2a. We applied a magnetic field of 6.02 T parallel to a axis in the tetragonal phase throughout the NMR measurements. In the nematic phase, the NMR spectra exhibit a double-edge structure \cite{Kuwa}, as shown in the left panel of Fig. 2a. The double edges have been observed as two separate peaks for pure FeSe \cite{Wang2016, Wiecki2017a, Baek2015, Wang2017c}. This edge structure disappears above 0.57 GPa. The spectra above 1 GPa exhibit a single peak. At 3.9 GPa, the $^{77}$Se signal disappears at approximately 60 K because of the AF order (see the right panel of Fig. 2a). The $T$ dependence of the linewidth at ambient pressure, 3.5 GPa and 3.9 GPa is shown in Fig. 2b. The AF order is observed via a remarkable increase in the linewidth and the loss of the signal. We defined $T_{\rm N}$ as the temperature of the onset of linewidth broadening.

\begin{figure*}
\includegraphics[clip, width=1\columnwidth]{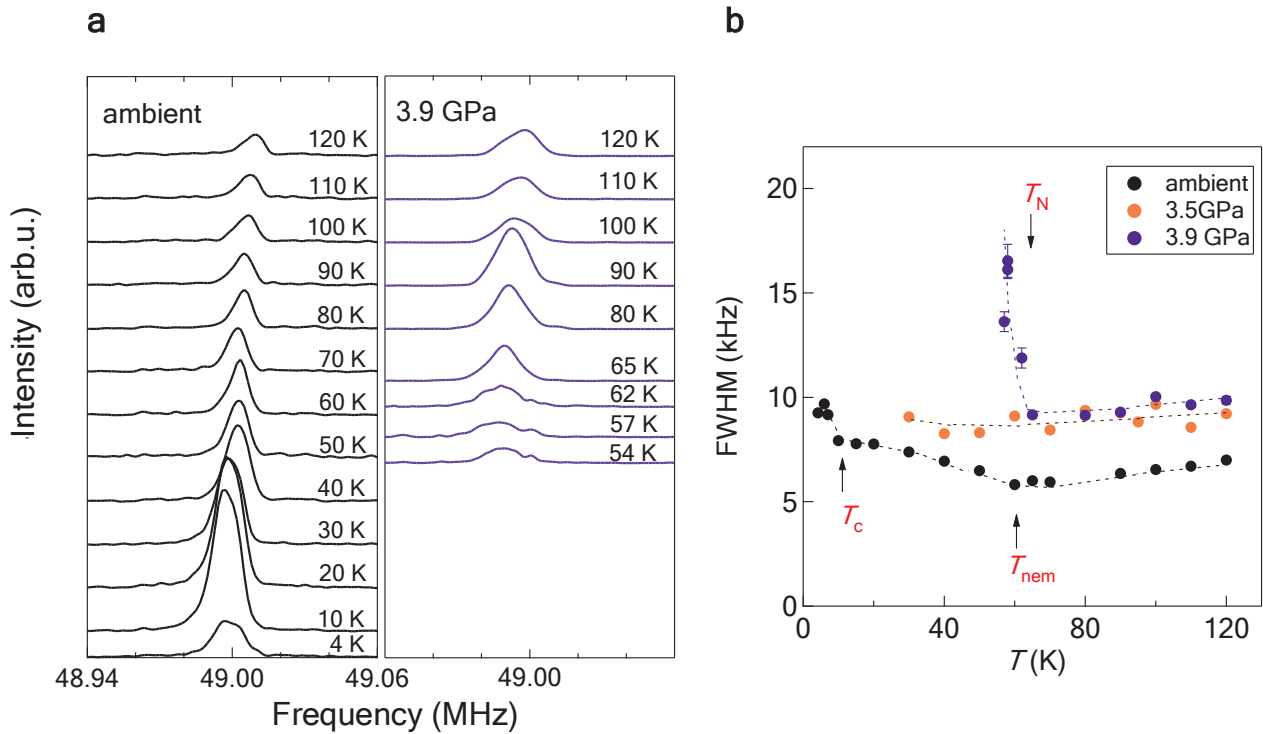}
\caption{\label{fig:epsart} { Typical $^{77}$Se-NMR spectra and linewidth for 12\% S-substituted FeSe corresponding to the nematic and antiferromagnetic (AF) orders.} {\bf a,} The NMR spectra at ambient pressure show a double-edge structure in the nematic phase below 60 K. The NMR spectra at 3.9 GPa broaden remarkably, and the signal is not observed below the AF transition temperature $T_{\rm N}$. {\bf b,} Full width at half maximum at 3.5 and 3.9 GPa. The arrows shown by $T_{\rm nem}$ and $T_{\rm c}$ represent the nematic and superconducting transition temperatures, respectively. }
\end{figure*}

Now, we focus on the Knight shift ($K$) in a paramagnetic state. Figure 3a shows the $T$ dependence of $K$ in a paramagnetic state. We adopted the average of the two edges for $K$ in the nematic phase. The data below 3 GPa were already publisged in an early work \cite{Kuwa}. The Knight shift above 3 GPa is $T$ dependent even at low temperatures suggesting the influence of AF fluctuations, whereas the influence is absent below 3 GPa. Hereafter, we discuss the $P$ dependence of $K$ below 3 GPa in relation to the DOS shown in Fig. 1b. Fig 3b shows the $P$ dependence of the NMR spectra at 60 K, and each spectrum
is fitted by a Gaussian function. From the peak positions in Fig. 3b, the $P$ dependence of $K$ is obtained, as shown in Fig. 3c. Note that the anomaly at 1 GPa is observed at entire temperatures and therefore is not directly caused by the nematic transition.

The Knight shift in a paramagnetic state is decomposed as
        \begin{equation}
	K=K_{\rm spin}+K_{\rm orb}
	\end{equation}
where $K_{\rm spin}$ and $K_{\rm orb}$ represent the spin and orbital parts of the Knight shift, respectively. The former and latter are $T$-dependent and $T$-independent, respectively. Experimentally, $K$ is decomposed into $K_{\rm spin}$ and $K_{\rm orb}$ using the uniform spin susceptibility, $\chi(0)$. The orbital part $K_{\rm orb}$ is estimated to be 0.26\% at ambient pressure [Supplemental material], which is almost the same as that obtained for pure FeSe \cite{JLi}. The results suggest that $K_{\rm orb}$ is insensitive to S substitution. The spin part $K_{\rm spin}$ and $\chi(0)$ are related to $K_{\rm spin}=A\chi(0)$, where $A$ is the hyperfine coupling. The monotonic decrease in $K$ below 3 GPa with decreasing $T$ suggests that the influence of magnetism is absent at low temperatures. In this case, $\chi(0)$ can be described using the formula for conventional paramagnetic metals and is related to the DOS of free electrons, $D(E_{\rm F})$. Therefore, $K_{\rm spin}$ is proportional to $D(E_{\rm F})$:
\begin{equation}
	K_{\rm spin} \propto D(E_{\rm F}).
	\end{equation}

Although the DOS shown in Fig. 1b is derived using a tight-binding (TB) model as described later, overall features can be roughly explained by the DOS for two-dimensional free electron systems. For two-dimensional free electron systems, $D(E_{\rm F})$ is expressed as
   \begin{equation}
	D(E_{\rm F})=\frac{(Na)^2}{2\pi} \frac{2m}{2\hbar}
	\end{equation}
where $N^2$, $a$, and $m$ are the total number of lattices, lattice constant, and electron mass, respectively. The $P$ dependence of $D(E_{\rm F})$ originates only from that of $a^2$. According to X-ray analyses up to 1 GPa \cite{Matsuura2017}, the lattice constant ($a$) shrinks linearly with increasing pressure. The lattice constant also shrinks for S substitution: 30\% S-substitution is equivalent to a pressure application of 1 GPa. Therefore, the discrepancy in $a$ between pure FeSe and 12\% S-substituted FeSe is trivial. We use $a$ for pure FeSe because $a$ for 12\% S-substituted FeSe is not available at present. In addition, the data above 1 GPa is not available at present, and instead we adopted the extrapolation of the data below 1 GPa. The values of $a^2$ and $K_{\rm spin}/a^2$ normalized by those at ambient pressure are shown in Fig. 3d. The normalized $K_{\rm spin}/a^2$ is a quantity compared with the theoretical results shown in Fig. 1b. The step-like enhancement at 1 GPa reaches 10\% of $K_{\rm spin}/a^2$ at ambient pressure, which is consistent with the theoretical calculation of the total DOS shown in Fig. 1b.  The enhancement of $K_{\rm spin}/a^2$ seems to be smaller than that shown in Fig. 1b, implying that the size of the hole pocket at the point $M$ is fairly small, as described later.  We
determined $K_{\rm spin}$ at low pressures below 1 GPa, assuming that $K_{\rm orb}$ is estimated to be $\sim 0.26 \%$. However, at high pressures such as 2 or 3 GPa, the determination of $K_{\rm spin}$ is very difficult because the data of $\chi(0)$ under pressure are not available. The assumption of $K_{\rm orb} \sim 0.26 \%$ leads to an unrealistic result, namely, $K_{\rm spin}$ or the DOS at high pressures becomes lower than that at ambient pressure. To overcome this difficulty, we focus on a remarkable drop in $K_{\rm spin}$ below $T_{\rm c}$ at 2 and 3 GPa (see Fig. 3a). The apparent drop at 2 and 3 GPa originates from $K_{\rm spin}$, indicating that $T$-independent $K_{\rm orb}$ decreases at high pressures. Therefore, we assumed that the decrease in $K_{\rm orb}$ at high pressures is equivalent to the drop in $K_{\rm spin}$ below $T_{\rm c}$. In Fig. 3d, we estimated the decrease in $K_{\rm orb}$ to be 0.005 and 0.01 \% for 2 and 3 GPa, respectively, from the drop in $K_{\rm spin}$ below $T_{\rm c}$.

\begin{figure*}
\includegraphics[clip, width=1\columnwidth]{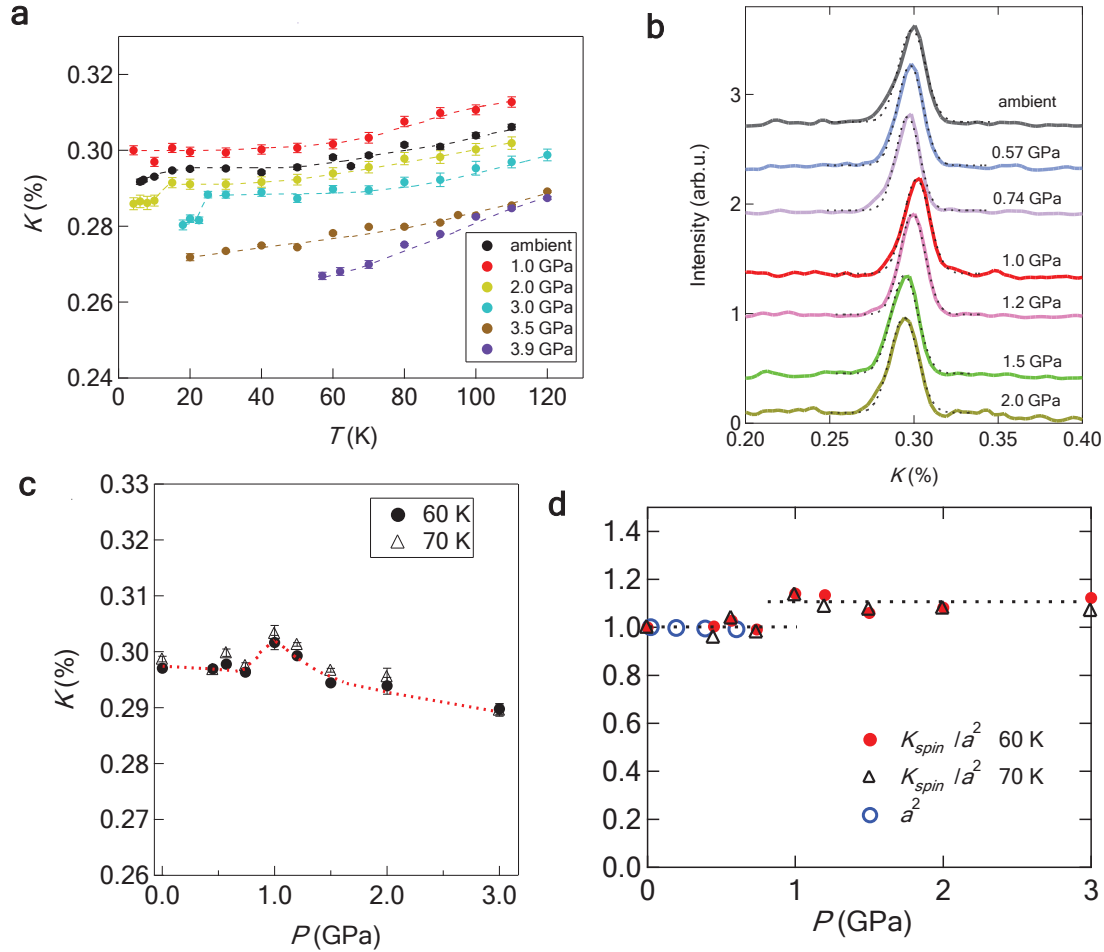}
\caption{\label{fig:epsart}  $^{77}$Se-Knight shift for 12\% S-substituted FeSe. {\bf a,} Temperature dependence of Knight shift ($K$) at several
pressures. The Knight shift below 3 GPa \cite{Kuwa} reflects the density of states (DOS) at low temperatures. The decrease in K due to
superconductivity is clearly observed at 2 or 3 GPa. {\bf b,} Pressure dependence of the NMR spectra at 60 K. Each spectrum is
fitted with a Gaussian function, as indicated by dotted curves. {\bf c,} Pressure dependence of $K$ measured at 60 and 70 K. The
dotted curves are guides to the eye. {\bf d,} Pressure dependence of the square of the a-axis lattice constant
($a^2$) and $K_{\rm spin}/a^2$ normalized by those at ambient pressure. The dashed lines represent the average of $K_{\rm spin}/a^2$ at pressure regions below and above 1 GPa.}
\end{figure*}

\begin{figure*}
\includegraphics [clip, width=1\columnwidth] {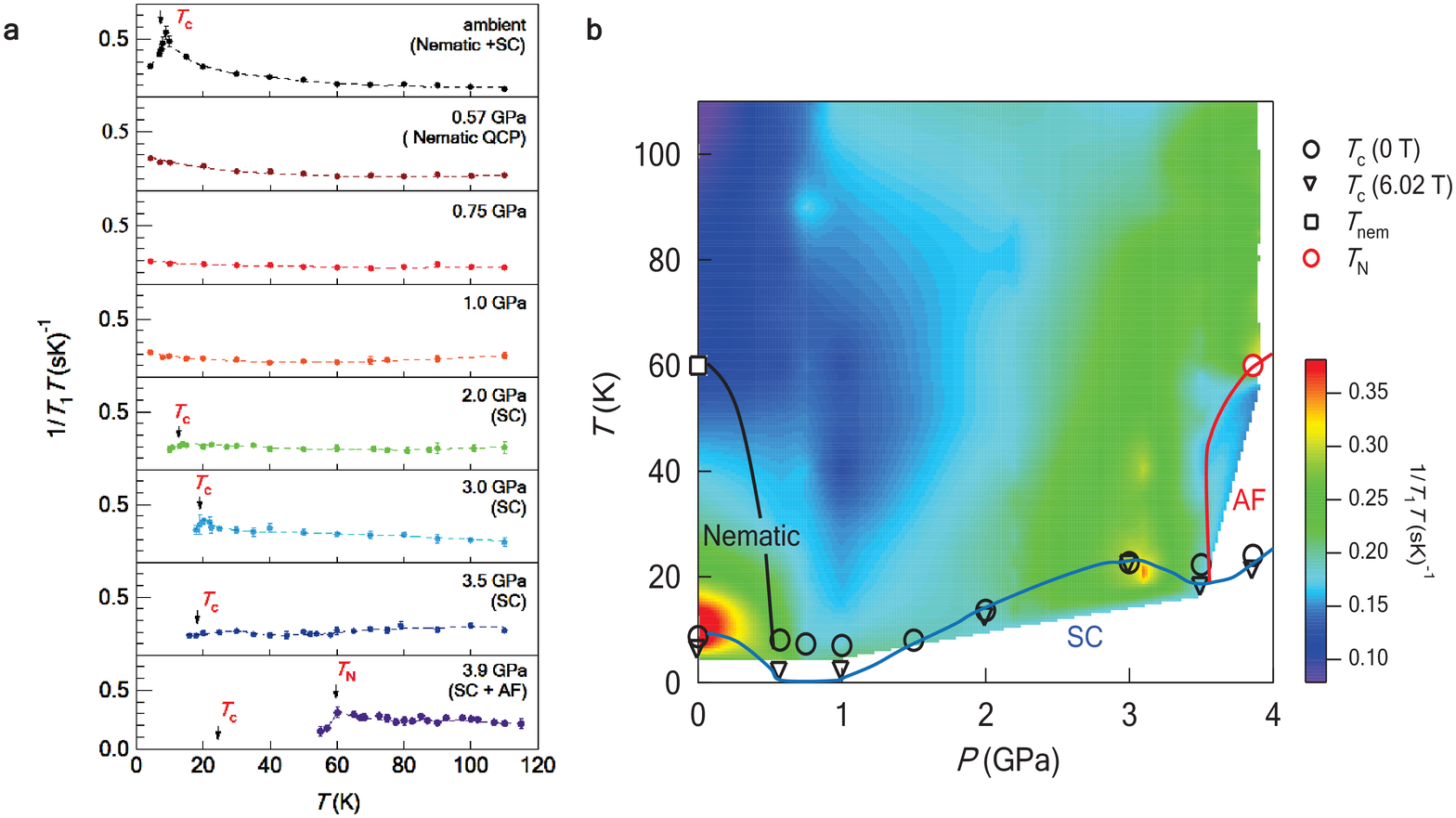}
\caption{\label{fig:wide}  $T$ dependence of $1/T_1T$ of $^{77}$Se measured at 6.02 T and color plot of $1/T_1T$. {\bf a,} The data at ambient pressure, 1.0, 2.0, and 3.0 GPa were published in an eraly work \cite{Kuwa}. $T_{\rm C}$s shown by arrows are determined from AC susceptibility measurements at 6.02 T. Both nematic and superconducting (SC) orders are absent at 1.0 GPa. {\bf b,} The SC phase shows a double-dome structure at 6.02 T. $T_{\rm C}$s shown by circles and inverted triangles are determined from AC susceptibility measurements at 0 T and 6.02 T, respectively. $T_{\rm C}$s shown by inverted triangles are the same as those shown by arrows in Fig. 4{\bf a}.}
\end{figure*}

The step-like enhancement is theoretically explained by the hole pocket at ($\pi$, $\pi$) appearing across the Fermi level owing to the lift of the $d_{xy}$ orbital (see Fig. 1a). Assuming that $a$=1, the $P$ dependence of the DOS for each orbital was calculated for 10\% S-substituted FeSe (Fig. 1b). The DOS was calculated using the TB models constructed from first-principles calculations based on the crystal structure. We denote the Hamiltonian for FeS$_x$Se$_{1-x}$ at pressure $P$ (GPa) as $H^{(0)}_{x}(P)$. The model Hamiltonian used for these calculations is expressed as

 \begin{equation}
	H(P)=H^{(0)}_{0.1}(0)+\Delta H(P) + \Delta E
	\end{equation}
where $\Delta E$ is the correction term added to fit the real size of the Fermi surfaces observed experimentally from ARPES and dHvA quantum oscillation \cite{Watson2015a, Watson2017, Kushnirenko2018, Coldea2018, Skornyakov2018, Fanfarillo2018}. Given that $\Delta H(P) \equiv H^{(0)}_{0.1}(P)-H^{(0)}_{0.1}(0)$ changes linearly in the pressure range of 0--4 GPa and is insensitive to the concentration $x$, $H^{(0)}_{0.1}(P) \simeq H^{(0)}_{0}(P)$, $\Delta H(P)$ is expressed as

 \begin{equation}
	\Delta H(P) \simeq \frac{P}{4} (H^{(0)}_{0}(4)-H^{(0)}_{0}(0)).
	\end{equation}
The DOS enhancement appears at around 1.5 GPa, which is consistent with the experimental results of $K$ shown in Fig. 2d.

The DOS enhancement can affect low-energy magnetic fluctuations obtained from the relaxation rates divided by temperature, $1/T_1T$, as the nesting condition between Fermi surfaces changes.
Figure 4a shows the $T$ dependence of $1/T_1T$ at several pressures. $T_{\rm c}$s shown by arrows were determined from the AC susceptibility measurements at 6.02 T [Supplemental material]. When the wave vector $(q)$-dependence of the hyperfine interaction is neglected, $1/T_1T$ is expressed as follows:
	\begin{equation}
	\frac{1}{T_1T} \propto \sum_q \frac{{\rm Im} \chi (q, \omega)}{\omega}
	\end{equation}
where $\omega$ and $\chi (q, \omega)$ represent the NMR frequency and the dynamical spin susceptibility, respectively.
Below $T_{\rm c}$, the signal intensity became extremely small, and thus, we could not detect signals below 10 K at 2.0 GPa and 15 K at 3.0 GPa, respectively. At ambient pressure, $1/T_1T$ shows Curie--Weiss-like behavior in the nematic phase, indicating the development of AF fluctuations \cite{Moriya1973, Moriya1974}:

\begin{equation}
	 \frac{1}{T_1T} \sim a + \frac{b}{T-\theta}
	\end{equation}
where $a$ and $b$ are independent of $T$. $\theta$ is estimated to be almost zero at ambient pressure. However, the Curie--Weiss behavior is strongly suppressed even at 0.57 GPa. Although the Curie--Weiss behavior is not clearly observed at pressures between 2 and 3.5 GPa, an anomaly of $1/T_1T$ is observed at around $T_{\rm c}$. In this pressure regime, $T_{\rm c}$ at 6.02 T gradually recovers with increasing pressure. The data at 3.9 GPa are completely different from those at 2.0 and 3.0 GPa in that the anomaly occurs not at $T_{\rm c}$ but at $T_{\rm N}$. The data series in Fig. 4a is presented as a color plot in Fig. 4b. As shown in Fig. 4b, different types of AF fluctuations are observed in the $P-T$ phase diagram, indicating that the origins of the AF fluctuations are different between the lower and higher pressure regimes. This result indicates a change in the nesting condition and confirms the theoretically predicted pressure-induced Lifshitz transition.

The results of $K$ and $1/T_1T$ show that the DOS and the AF fluctuations change at around 1 GPa, respectively. Interestingly, the AF fluctuations, which are unambiguous in the low-pressure regime where the nematic order occurs, unexpectedly become ambiguous in the high-pressure regime despite the AF phase boundary. In general, the Curie--Weiss behavior should be clearly observed near the AF phase boundary.
Therefore, ambiguous AF fluctuations at high pressures are extremely rare compared to those of conventional AF magnets. This peculiarity arises because the nesting condition is not optimal, which implies
that the $d_{xy}$ hole pocket is fairly small. Such a small hole
pocket is consistent with the small increase in the
DOS at 1 GPa.

Another peculiar phenomenon is the loss of the NMR signal at low temperatures in the high-pressure regime above 1 GPa.
This peculiarity can be attributed to the close relationship between the nematic and AF orders \cite{Yamakawa2017a}. Nematic and/or AF states can appear at any pressure. Thus, a short-range AF order can develop even in the absence of a long-range AF order. A short-range AF order would make the NMR signal very weak and undetectable at low $T$ below $T_{\rm c}$. A long-range AF order at pressures above 3.9 GPa can develop together with a finite order parameter from the short-range AF order.

We demonstrated the strong suppression of $1/T_1T$ under pressure.
A similar suppression is also observed by isovalent S substitution \cite{Wiecki2018}, despite the fact that the Fermi surfaces become larger and the nesting condition becomes better with increasing S concentration \cite {Coldea2016, Watson2015b, Reiss2017}. S substitution would have the same effect as the application of pressure, because the atomic radius of S is smaller than that of Se.
However, the chalcogen height decreases with increasing S concentration, in contrast to the application of pressure \cite{Matsuura2017}.
For the heavily S-substituted regime over 20\%, where the BCS-BEC crossover has been suggested \cite{Kasahara2014, Hosoi2016, Sato2017, Hanaguri2017, Hanaguri2019}, the Curie--Weiss behavior of $1/T_1T$ is strongly suppressed, similar to $1/T_1T$ for 12\% S-substituted FeSe at 1 GPa.
Although the strong suppression of the Curie--Weiss behavior is common, it is not clear whether the present high-pressure regime is smoothly linked with the heavily S-substituted regime.
To solve this problem, further investigation is needed.

In conclusion, we performed $^{77}$Se-NMR measurements on 12\% S-substituted FeSe under pressures of up to 3.9 GPa. We observed the anomalies of $K$ and $1/T_1T$ corresponding to the theoretically predicted pressure-induced Lifshitz transition. These results indicate that nematicity and magnetism exhibit cooperative coupling. The AF fluctuation unambiguously develops in the nematic phase as the Curie-Weiss behavior of $1/T_1T$ in the low-pressure regime below 1 GPa. In contrast, in the high-pressure regime between 1 and 3.9 GPa where the nematic order is absent, the AF fluctuation is strongly suppressed. Further, a high $T_{\rm c}$ is realized in a pressure regime where the nematic order is absent and the correlated AF fluctuation is fairly weak. The emergence of the $d_{xy}$ orbital and its intra-orbital coupling play a key role for the high-$T_{\rm c}$ superconductivity.

\section*{Methods}

We performed $^{77}$Se-NMR measurements at 6.02 T using a single crystal of 12\% S-substituted FeSe with dimensions of approximately $1.0 mm \times 1.0 mm \times 0.5 mm$.
We applied a magnetic field parallel to the FeSe planes to suppress the decrease in $T_{\rm c}$.
We applied a pressure up to 3.9 GPa using a NiCrAl piston-cylinder-type pressure cell. The highest pressure attainable by clamping this pressure cell is 3.7 GPa because a decrease of 10\% in pressure is inevitable after releasing a load. To attain a pressure of 3.9 GPa, we maintained a constant load by employing an oil press mounted on top of the cryostat \cite{Fujiwara2007}.
We performed pulsed-NMR measurements using a conventional spectrometer and measured the relaxation time ($T_1$) via the saturation-recovery method.

\bibliography{References}

\section*{Data availability}
Data are available from the corresponding author upon reasonable request.

\section*{Acknowledgements (not compulsory)}

The present work was supported by Grants-in-Aid for Scientific Research (KAKENHI Grant No. JP18H01181) and a grant from the Mitsubishi Foundation. This work was partly supported by Grants-in-Aid for Scientific Research (KAKENHI Grant Nos. JP18H05227, JP19H00648, and JP18K13492) and by Innovative Areas “Quantum Liquid Crystals” (No. JP19H05824) from the Japan Society for the Promotion of Science.

\section*{Author contributions statement}

N. F. designed the NMR experiments. T. K. carried out the NMR measurements under the instruction of N. F.. K. M., Y. M., S. K., Y. M., and T. S. synthesized the samples and performed the chemical analysis of the samples. J. G. measured the magnetization at ambient pressure to decompose the Knight shift to the orbital and spin parts. Y. Y. and H. K. calculated the DOS. The high-pressure technique under a constant load was designed by Y. U..

\newpage

\bf{Supplemental Materials} 

\begin{enumerate}
\item Determination of $T_{\rm c}$ from the AC susceptibility
\item Magnetization
\item Orbital part of the Knight shift determined from the K-$\chi$ plot

\end{enumerate}

\section {Determination of $T_{\rm c}$ from the AC susceptibility}

We determined $T_{\rm c}$ from the resonance frequency $f_{\rm{r}}$ of the tank circuit attached to the head of an NMR probe. The frequency $f_{\rm{r}}$ was measured using a commercially available network analyzer. It is related to the AC susceptibility $\chi$ as $f_{\rm{r}}=1/\sqrt{LC(1+4\pi\chi)}$, where $C$ and $L$ represent the capacitance of a variable capacitor and the inductance of a coil wound onto the sample, respectively. Figure 1 shows the $T$ dependence of $f_{\rm{r}}$: $f_{\rm{r}}$ increases gradually with decreasing temperature from room temperature, as $L$ of the coil gradually decreases during the cooling process. The drastic increase in $f_{\rm{r}}$ occurs at $T_{\rm c}$ owing to the Meissner effect. We determined $T_{\rm c}$s from the crossing points of the dashed lines, as indicated by the red arrows in Fig. 1. $T_{\rm c}$ at ambient pressure is approximately 9 K. This result is in agreement with $T_{\rm c}$ determined from magnetization measurements, as described below.

\begin{figure*}[b]
\includegraphics [clip, width=0.7\columnwidth] {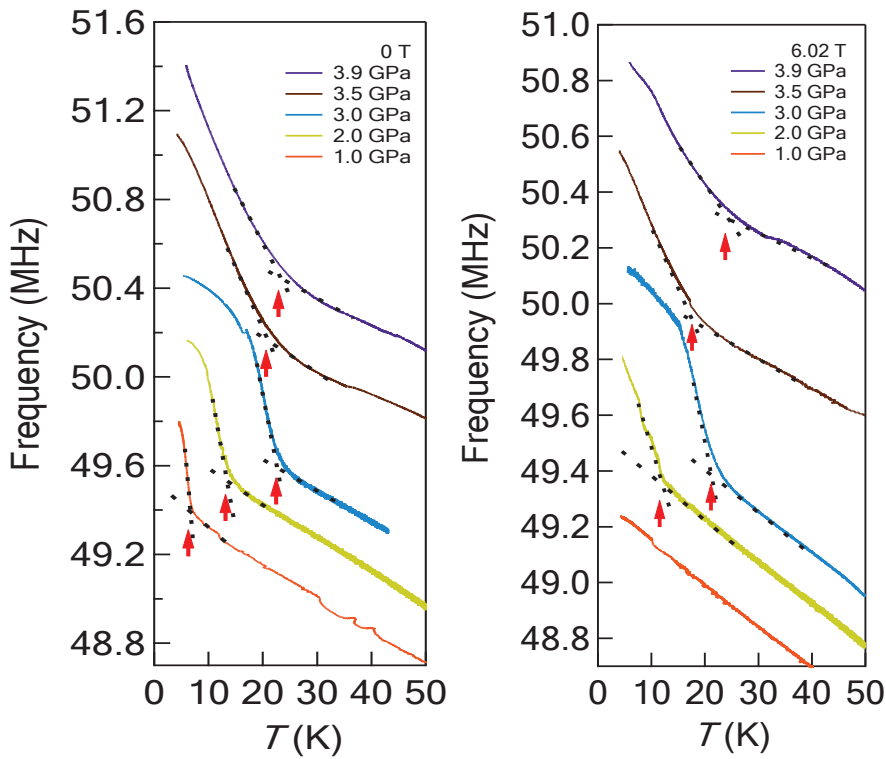}
\caption{\label{fig:wide} $T$ dependence of the resonance frequency $f_{\rm{r}}$ of the tank circuit measured at zero and 6.02 T. We measured $f_{\rm{r}}$ at 6.02 T because the NMR measurements described in the main text were performed at 6.02 T. }
\end{figure*}

\section {Magnetization and DC susceptibility}

Figure 2 shows the $T$ dependence of the magnetization ($M$) and susceptibility ($\chi$) measured for 12\% S-substituted FeSe with an MPMS system by Quantum Design Inc. $T_{\rm c}$ determined from the magnetization is in good agreement with that determined from the AC susceptibility. A large demagnetization was observed at $T_{\rm c} \sim 9 K$. The susceptibility data measured at 6 T were used to separate the spin and orbital parts of the Knight shift, as shown in the following $K-\chi$ plot.

\begin{figure*}
\includegraphics [clip, width=1\columnwidth] {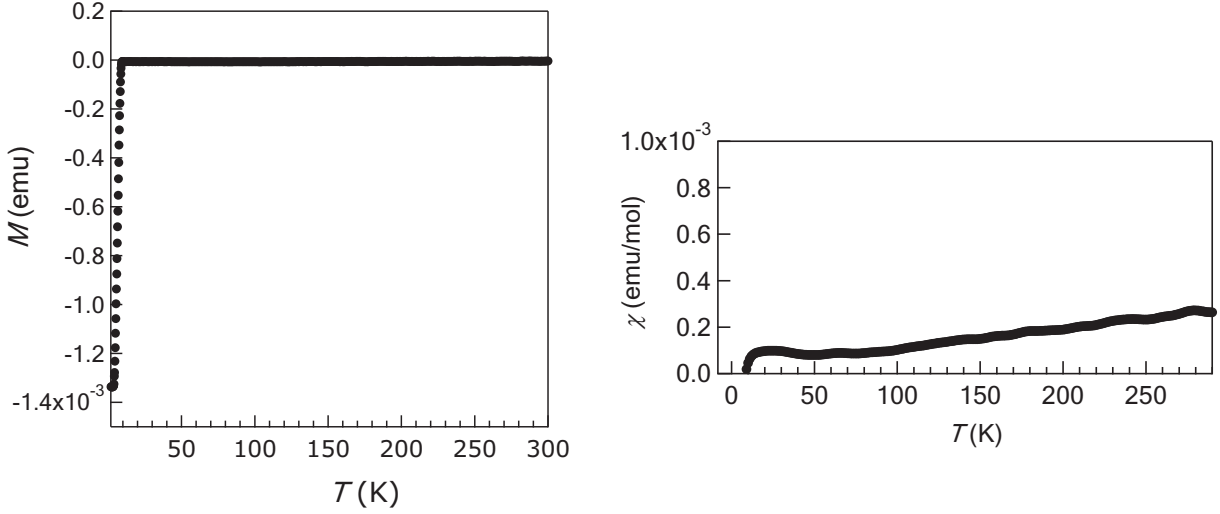}
\caption{\label{fig:wide} Magnetization and susceptibility for 12\% S-substituted FeSe }
\end{figure*}

\section {Spin and orbital parts of $^{77}K$ determined from the $^{77}K-\chi$ plot}

The susceptibility $\chi$ measured from the experiments is decomposed into three parts: the uniform spin part $\chi(0)$, orbital part $\chi_{\rm orb}$, and diamagnetic term $\chi_{\rm dia}$; $\chi=\chi(0)+\chi_{\rm orb}+\chi_{\rm dia}$. The last term is estimated to be approximately $6\times10^{-5}$ (emu/mol Oe). Both $\chi(0)$ and $\chi_{\rm orb}$ originate from $d$ electrons, and they are estimated only from the $^{57}K-\chi$ plot for $^{57}$Fe. $\chi(0)$ and $\chi_{\rm orb}$ are expressed as $^{57}K_{\rm spin}=^{57}A\chi(0)$ and $^{57}K_{\rm orb}=^{57}A_{\rm orb}\chi_{\rm orb}$, where $^{57}K_{\rm spin}$ and $^{57}K_{\rm orb}$ represent the spin and orbital parts of $^{57}K$, respectively.  $^{57}A$ is the hyperfine coupling of $^{57}$Fe and is obtained from the slope of the $^{57}K-\chi$ plot. $^{57}A_{\rm orb}$ is given as $2<\frac{1}{r^{3}}>$, where $r$ represents the distance between an electron on the $d$ orbits and the $^{57}$Fe nucleus. Because five $d$ orbitals are involved in iron-based systems, detailed theoretical investigations are needed to restrict the estimation of $^{57}A_{\rm orb}$. Because such theoretical calculations are currently unavailable, $^{57}A_{\rm orb}$ is estimated, for convenience, using $<\frac{1}{r^{3}}>$ for free Fe$^{3+}$ ions. Five parameters $\chi(0), \chi_{\rm orb}, \chi_{\rm dia}, ^{57}K_{\rm spin}$, and $^{57}K_{\rm orb}$ are uniquely determined from the following five equations:

\begin{eqnarray}
^{57}K=^{57}K_{\rm orb}+^{57}K_{\rm spin} \\
\chi=\chi(0)+\chi_{\rm orb}+\chi_{\rm dia} \\
^{57}K_{\rm spin}=^{57}A\chi(0) \\
^{57}K_{\rm orb}=^{57}A_{\rm orb}\chi_{\rm orb} \\
\chi_{\rm dia} \sim 6\times10^{-5}
\end{eqnarray}
where $^{57}$A is determined as the slope of the $^{57}K-\chi$ plot and $^{57}A_{\rm orb}$ is calculated using the parameter of free Fe$^{3+}$ ions. Li {\it et al.} estimated $\chi_{\rm orb}$ based on the assumption that $\chi_{\rm orb}+\chi_{\rm dia} \sim 0$, namely $\chi \sim \chi(0)$ instead of $^{57}K_{\rm orb}=^{57}A_{\rm orb}\chi_{\rm orb}$ [1].

The Knight shift ($^{77}K$) of $^{77}$Se is obtained from the $^{77}$Se-NMR spectra. $^{77}K$ is decomposed into the orbital part ($^{77}K_{\rm orb}$) and the spin part ($^{77}K_{\rm spin}$): $^{77}K=^{77}K_{\rm orb}+^{77}K_{\rm spin}$. The latter is related to the uniform spin susceptibility $\chi(0)$; therefore, the density of states ($^{77}K_{\rm spin}=^{77}A\chi(0)$, where $^{77}A$ is the hyperfine coupling of $^{77}$Se). The $^{77}K-\chi$ plot is often used for the decomposition of $^{77}K$ into $^{77}K_{\rm orb}$ and $^{77}K_{\rm spin}$, as shown in Fig. 3.
Li {\it et al.} also estimated $^{77}K_{\rm orb}$ to be 0.23\% for pure FeSe from the $^{77}K-\chi$ plot under the assumption that $\chi_{\rm orb}+\chi_{\rm dia} \sim 0$. The $^{77}K-\chi$ plot for 12\% S-substituted FeSe is almost the same as that for pure FeSe [1]; therefore, $^{77}K_{\rm orb}$ can be estimated to be 0.26\%, similar to the case of pure FeSe. \ \\

\begin{figure*}
\includegraphics [clip, width=0.5\columnwidth] {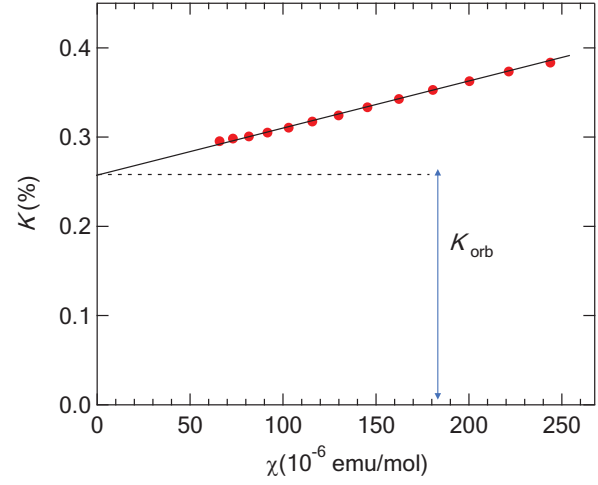}
\caption{\label{fig:wide} $^{77}K-\chi$ plot for 12\% S-substituted FeSe }
\end{figure*}

[1] J. Li, B. Lei, D. Zhao, L. P. Nie, D. W. Song, L. X. Zheng, S. J. Li, B. L. Kang, X. G. Luo, T. Wu, and X. H. Chen, Supplemental material, Phys. Rev. X 10, 011034 (2020).
\end{document}